\documentclass[10pt,conference]{IEEEtran}
\IEEEoverridecommandlockouts
\usepackage{amsmath,amssymb,amsfonts}
\usepackage{paralist}
\usepackage{algorithmic}
\usepackage{graphicx}
\usepackage{textcomp}
\usepackage{tabularx}
\usepackage{xcolor}
\usepackage{xspace}
\usepackage{hyperref}
\usepackage{balance}
\usepackage{blindtext}
\usepackage{booktabs}

\usepackage{cite}

\usepackage{soul}
\usepackage{listings}
\def\BibTeX{{\rm B\kern-.05em{\sc i\kern-.025em b}\kern-.08em
    T\kern-.1667em\lower.7ex\hbox{E}\kern-.125emX}}

\usepackage{url}

\definecolor{codegreen}{rgb}{0,0.6,0}
\definecolor{codegray}{rgb}{0.5,0.5,0.5}
\definecolor{codepurple}{rgb}{0.58,0,0.82}
\definecolor{backcolour}{rgb}{0.95,0.95,0.92}
\lstdefinestyle{mystyle}{
    backgroundcolor=\color{backcolour},   
    commentstyle=\color{codegreen},
    keywordstyle=\color{magenta},
    numberstyle=\tiny\color{codegray},
    stringstyle=\color{codepurple},
    basicstyle=\ttfamily\footnotesize,
    breakatwhitespace=false,         
    breaklines=true,                 
    captionpos=b,                    
    keepspaces=true,                 
    numbers=left,                    
    numbersep=5pt,                  
    showspaces=false,                
    showstringspaces=false,
    showtabs=false,                  
    tabsize=2
}
\newcommand{\tool}[0]{\textsc{TEASER}\xspace}
\newcommand{\scissor}[0]{\textsc{SDC-Scissor}\xspace}
\newcommand{\beamng}[0]{BeamNG.tech\xspace}
\newcommand{\carla}[0]{CARLA\xspace}
\newcommand{\aicas}[0]{\textit{aicas GmbH}\xspace}
\newcommand{\pythoncan}[0]{\texttt{python-can}\xspace}
\newcommand{\cantools}[0]{\texttt{cantools}\xspace}

\title{\tool: Simulation-based CAN Bus Regression Testing for Self-driving Cars Software}

\author{Christian Birchler$^{\ast, \dag}$, Cyrill Rohrbach$^{\dag}$, Hyeongkyun Kim$^{\ddag}$, \\ Alessio Gambi$^{\S}$, Tianhai Liu$^{\|}$, Jens Horneber$^{\|}$, Timo Kehrer$^{\dag}$, Sebastiano Panichella$^{\ast}$ \\
$\ast$ \textit{Zurich University of Applied Sciences, Switzerland} \\
$\dag$ \textit{University of Bern, Switzerland} \\
$\ddag$ \textit{University of Zurich, Switzerland} \\
$\S$ \textit{IMC University of Applied Sciences Krems, Austria} \\
$\|$ \textit{aicas GmbH, Germany}
}

\begin{document}
\maketitle

\begin{abstract}
Software systems for safety-critical systems like self-driving cars (SDCs) need to be tested rigorously.
Especially electronic control units (ECUs) of SDCs should be tested with realistic input data.
In this context, a communication protocol called Controller Area Network (CAN) is typically used to transfer sensor data to the SDC control units.
A challenge for SDC maintainers and testers is the need to manually define the CAN inputs that realistically represent the state of the SDC in the real world.
To address this challenge, we developed \tool,
which is a tool that generates realistic CAN signals for SDCs obtained from sensors from state-of-the-art car simulators.
We evaluated \tool based on its integration capability into a DevOps pipeline of \aicas, a company in the automotive sector.
Concretely, we integrated \tool in a Continous Integration (CI)  pipeline configured with Jenkins.
The pipeline executes the test cases in simulation environments and sends the sensor data over the CAN bus to a physical CAN device, which is the test subject.
Our evaluation shows the ability of \tool to generate and execute CI test cases that expose simulation-based faults (using regression strategies);
the tool produces CAN inputs that realistically represent the state of the SDC in the real world.
This result is of critical importance for increasing automation and effectiveness of simulation-based CAN bus regression testing for SDC software.

\noindent \textit{Tool}: \url{https://doi.org/10.5281/zenodo.7964890} \\ \textit{GitHub}: \url{https://github.com/christianbirchler-org/sdc-scissor/releases/tag/v2.2.0-rc.1} \\ \textit{Documentation}: \url{https://sdc-scissor.readthedocs.io}
\end{abstract}


\section{Introduction}
In recent years, the deployments of autonomous systems such as self-driving cars (SDCs) and unmanned aerial vehicles, several accidents happened~\cite{us-adas-accidents,NPR,9news,KhatiriPT23,UAVtosem}.
Hence, those incidents imply the importance of testing for safety-critical systems such as SDCs.
Using simulation environments to test SDCs brings several advantages over real-world testing in the field, especially the aspects of reproducibility, safety, and determinism of the test cases.
However, testing in simulation is costly in terms of computational power and time; therefore, it is required to do it effectively.
Testing on the system level of SDCs focuses on the correct interaction of different components of the vehicle, such as the engine control module, transmission control module, brake control module, etc.
Those components, also known as electronic control units (ECUs), interact with each other with a common protocol.
In the automotive domain, the CAN bus protocol is a widely used communication standard for ECUs~\cite{CiA}.
CAN bus allows the communication of different ECUs in a vehicle over a shared bus system by a standardized protocol~\cite{can-iso-standard}.
The main challenge for SDC maintainers and testers is to generate realistic test CAN signals which accurately reflect the state of an SDC in the real world since CAN signals are still manually generated for testing purposes nowadays (e.g., at \aicas).

The research on testing with CAN bus focuses on security, model-based testing~\cite{DBLP:conf/IEEEares/Huang0B18,DBLP:conf/ACISicis/YangTS14,DBLP:conf/adhocnets/CrosTC20,8658720} and CAN queuing~\cite{DBLP:conf/bic-ta/ZhangL13a}.
The research on CAN signals generation based on simulation environments, however, was mainly outside of the SDC domain~\cite{DBLP:conf/itsc/VershininNS15,7061002,DBLP:journals/jcm/LiuL13a}.
Hence, to the best of our knowledge, there is no tool that supports regression testing for SDC software on ECUs based on the CAN bus protocol with realistic input data.
We aim to do simulation-based regression testing for self-driving cars with their ECUs by using different simulators and the vehicles' CAN bus system.

To enable research on this problem, we developed \tool (simula\textbf{T}ion bas\textbf{E}d c\textbf{A}n bu\textbf{S} t\textbf{E}sting), a tool for simulation-based CAN bus testing that translates simulated sensor data of an SDC, obtained from a simulation environment, for the CAN bus transmission.
We conjecture the use of sensor data from multiple different simulation environments produces more realistic CAN signals for testing, which helps to detect software defects of ECUs.
Furthermore, \tool mitigates the currently common practice of manually generating CAN signals to test ECUs (as done by \aicas).

The contribution of this paper is threefold:
\begin{inparaenum}[(i)]
    \item \tool is publicly available on GitHub as a feature component of \scissor~\cite{9825849} with a GPLv3~\footnote{\url{https://www.gnu.org/licenses/gpl-3.0}} license.
    \item \tool reduces the time for generating realistic CAN bus signals for testing CAN devices, as demonstrated at \aicas
    \item we qualitatively evaluated the usefulness of \tool in the industrial setting of \aicas by integrating it into their DevOps pipeline for testing a physical CAN device.
\end{inparaenum}


\section{The \tool tool}\label{sec:tool}

\subsection{Architecture overview and main scenarios}
\begin{figure}
    \centering
    \includegraphics[width=\linewidth]{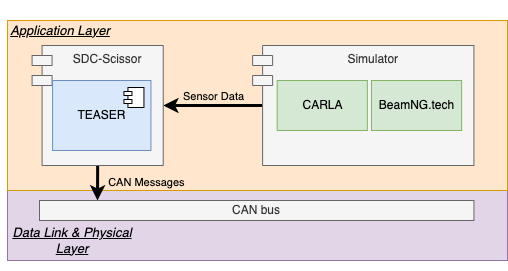} 
    \caption{\tool system view}
    \label{fig:system-view}
\end{figure}

The high-level architecture of the system is illustrated in Figure~\ref{fig:system-view}.
\tool is fully integrated as a component in \scissor~\cite{9825849}, which is a tool that uses machine learning to select simulation-based test cases for SDCs, and extends the existing tool with CAN bus functionalities.
With \tool we can generate CAN signals based on simulated scenarios from different simulators such as \beamng and \carla.
Therefore, \tool enables regression testing based on CAN signals which are now realistically simulated by the virtual environments.

\subsection{Simulation environments: \beamng and \carla}
\tool supports two simulation environments to generate CAN signal from.
The first simulator is the \beamng simulator.
\beamng simulates soft-body physics behavior in its virtual environment.
The second simulator is \carla.
In contrast to the \beamng simulator, \carla simulates a rigid-body physics behavior.
Both simulators are widely used in academia and in practice~\cite{DBLP:conf/sbst/2022,DBLP:conf/sbst/2021,birchler2023machine}.

\subsection{Approach and technological overview}

\begin{figure*}[t]
    \centering
    \includegraphics[width=\textwidth]{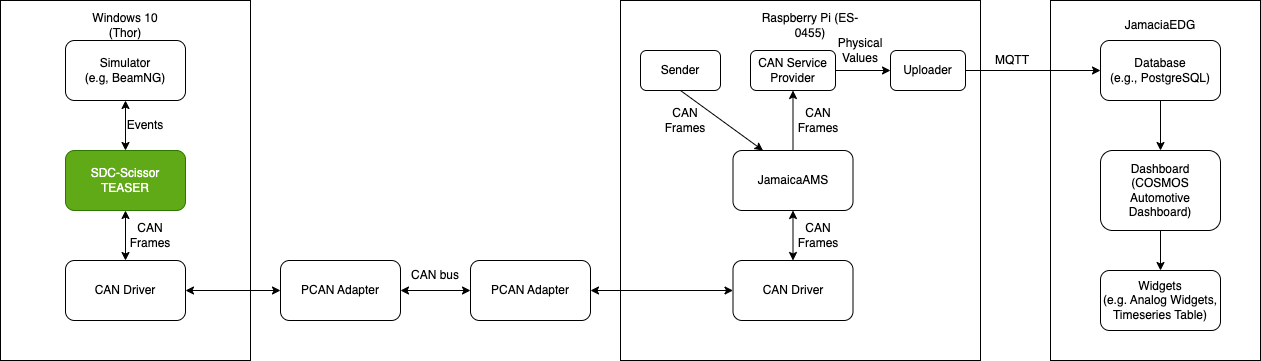} 
    \caption{Infrastructure at \aicas}
    \label{fig:aicas-infrastructure}
\end{figure*}

\tool's main objective is to extend the test runner of \scissor to enable CAN bus testing.
The tool uses two open source python libraries; the \pythoncan~\footnote{\url{https://github.com/hardbyte/python-can}} and \cantools~\footnote{\url{https://github.com/cantools/cantools}} packages.
The \pythoncan library allows communication with the CAN bus over specific interfaces (e.g., sockets).
Complementary to the first  package, the \cantools library provides functionality to compose the can messages to send on the CAN bus.
Specifically, \cantools allows the user to specify a CAN database file, which defines how signals are encoded into CAN messages.
The Listing~\ref{lst:can-dbc} illustrates how the wheel speed, throttle, brake, and steering angle are encoded in a CAN message by specifying it in a CAN database file.

\begin{lstlisting}[caption={Sample entries of a CAN database file}, label={lst:can-dbc}]
...

BO_ 177 sampleFrame2: 4 Vector__XXX
 SG_ wheelspeed : 16|16@1+ (0.2,0) [0|13107] "rpm" Vector__XXX

BO_ 161 sampleFrame1: 7 Vector__XXX
 SG_ throttle : 16|16@1+ (0.0001,0) [0|1] "%" Vector__XXX
 SG_ brake : 0|16@1+ (0.0001,0) [0|1] "%" Vector__XXX
 SG_ steering : 32|17@1- (0.01,0) [-655.36|655.35] "degree" Vector__XXX

 ...
\end{lstlisting}

In a nutshell, all implementations were done in the context of the \texttt{label-tests} subcommand of \scissor.
The input number of arguments for the subcommand is increased by CAN-specific information as illustrated in Listing~\ref{lst:can-args}.

\begin{lstlisting}[label={lst:can-args}, caption={Configuration file with highlighted CAN arguments}]
command: 'label-tests'
options:
  home: 'C:\BeamNG.tech.v0.24.0.2\BeamNG.drive-0.24.0.2.13392'
  user: 'C:\BeamNG.drive'
  tests: 'C:\Users\birch\repositories\sdc-scissor\destination'
  rf: 1.5
  oob: 0.3
  max_speed: 50
  interrupt: false
  obstacles: false
  bump_dist: 20
  delineator_dist: null
  tree_dist: null
  field_of_view: 120
  (*@\hl{canbus: true}@*)
  (*@\hl{can\_stdout: true}@*)
  (*@\hl{can\_dbc: '/path/to/beamng\_pipeline\_sample.dbc'}@*)
  (*@\hl{can\_dbc\_map: '/path/to/dbc\_map\_beamng.json'}@*)
  (*@\hl{can\_interface: 'socketcan'}@*)
  (*@\hl{can\_channel: 'vcan0'}@*)
  (*@\hl{can\_bitrate: 250000}@*)
  (*@\hl{influxdb\_bucket: 'your\_InfluxDB\_bucket'}@*)
  (*@\hl{influxdb\_org: 'your\_InfluxDB\_organization'}@*)
\end{lstlisting}

\section{Using \tool tool}\label{sec:usage}
\subsection{Requirements}
The following external software systems are required:
\begin{inparaenum}[(i)]
    \item Windows 10
    \item Python 3.10
    \item Pip
    \item \beamng~\footnote{\url{https://beamng.tech/}} v0.24 and/or \carla~\footnote{\url{https://carla.org/}}
    \item Poetry~\footnote{\url{https://python-poetry.org/}} (optional), and
    \item InfluxDB~\footnote{\url{https://www.influxdata.com/}} (optional)
\end{inparaenum}

The following instructions assume a full installation of the mentioned requirements on a Windows 10 machine.

\subsection{Instructions}
Get \scissor with the \tool directly from Zenodo~\footnote{\url{https://doi.org/10.5281/zenodo.7964890}}, GitHub~\footnote{\url{https://github.com/christianbirchler-org/sdc-scissor/releases/tag/v2.2.0-rc.1}} or PyPI~\footnote{\url{https://pypi.org/project/sdc-scissor/2.2.0rc1/}} and run the \tool component by invoking the \texttt{label-tests} subcommand.
For more details, also consolidate the demonstration video~\footnote{\url{https://doi.org/10.5281/zenodo.7965263}}.
\begin{lstlisting}[]
git clone https://github.com/christianbirchler-org/sdc-scissor.git
cd sdc-scissor
poetry install
poetry run sdc-scissor label-tests [args...]
\end{lstlisting}

An overview of all subcommands with their options is provided when invoking the \texttt{--help} flag.
\begin{lstlisting}[]
poetry run sdc-scissor label-tests --help
...
\end{lstlisting}

Specifying the commands and their options can also be done inside a configuration file as illustrated in Listing~\ref{lst:can-args}.
To invoke \tool with the configuration file the \texttt{-c} option is provided:
\begin{lstlisting}[]
poetry run sdc-scissor -c /path/to/config.yml
\end{lstlisting}

\tool extends the existing argument options; we need to use the highlighted arguments in Listing~\ref{lst:can-args}.
Table~\ref{tab:teaser-arguments} is an overview of the arguments with their according data type and description.

\begin{table}[t]
    \centering
    \caption{\tool arguments}
    \scriptsize
    \begin{tabularx}{\linewidth}{llX}
        \toprule
        \textbf{Argument} & \textbf{Type} & \textbf{Description} \\
        \midrule
        \texttt{canbus} & Boolean & Indicator if \tool should be enabled \\
        \midrule
        \texttt{can\_stdout} & Boolean & Indicate if \tool should print the CAN messages to \texttt{stdout}, i.e., to the console \\
        \midrule
        \texttt{can\_dbc} & String & Path to the CAN database file, which consists of data encoding information \\
        \midrule
        \texttt{can\_dbc\_map} & String & Path to a DBC map file, which consists of information on how to assess the data from different simulators \\
        \midrule
        \texttt{can\_interface} & String & Specifying the interface to use for CAN \\
        \midrule
        \texttt{can\_channel} & Strick & Specifingy the channel to use for CAN \\
        \midrule
        \texttt{can\_bitrate} & Integer & Bitrate to have for CAN \\
        \midrule
        \texttt{influxdb\_bucket} & String & Bucket name of an InfluxDB instance to use or to create if it does not exist yet \\
        \midrule
        \texttt{influxdb\_org} & String & The organization of the InfluxDB to use \\
        \bottomrule
    \end{tabularx}
    \label{tab:teaser-arguments}
\end{table}

If an InfluxDB instance is in use, then the respective API access token and host must be specified as environment variables.
\tool provides the option to declare the environment variable in a \texttt{.env} file:
\begin{lstlisting}[]
INFLUXDB_TOKEN="SeCrEtToKeN"
INFLUXDB_URL="http{s}://influxdb.example.org:{PORT}"
\end{lstlisting}
Alternatively, the environment variables can be set explicitly from the Windows control panel.

The \tool component provides different options to output the CAN messages:
\begin{inparaenum}[i)]
    \item Standard output (\texttt{stdout})
    \item Physical CAN interface defined in the configuration
    \item dumping the signals to an InfluxDB instance, or
    \item any combination of the previous possibilities.
\end{inparaenum}
This output behavior is achieved through implementing them by applying the decorator design pattern.

\section{Evaluation}\label{sec:evaluation}
As demonstrated in~\cite{birchler2023machine}, \scissor achieves an accuracy of 70\%, a precision of 65\%, and a recall of 80\% in selecting tests leading to a fault and improved testing cost-effectiveness.
The usefulness of \scissor with \tool in an industrial context is also demonstrated and explained~\cite{birchler2023machine}, where a tester at \aicas requires two days to produce CAN signals manually for 15 test cases.
The automation with \tool significantly reduces the time to generate realistic CAN signals since they are generated at runtime, where on average, a single test case in simulation with \beamng requires 49 seconds~\cite{10.1145/3533818}.

Furthermore, the video~\footnote{\url{https://doi.org/10.5281/zenodo.7964959}} shows the integration at \aicas use-case whose infrastructure is illustrated in Figure~\ref{fig:aicas-infrastructure}.
At \aicas we have a simulation environment installed on a Windows machine.
The simulation starts when a build job from Jenkins is triggered.
When the simulation starts, the \tool component produces the CAN frames and sends them over the CAN bus.
A Raspberry Pi, which represents a physical CAN device, receives the messages.
A separate application (JamaicaEDG) connects to the CAN device and displays on a dashboard the transmitted values from the CAN bus.
Specifically, the speed and throttle values are represented.

\section{Implications \& Future work}\label{sec:discussion-conclusion}

With \tool, regression testing for SDCs is not limited towards the fully black box approach as initially done by~\cite{9825849,10.1145/3533818,7173936,birchler2023machine} by neglecting ECU components and their interactions;
instead, testing of individual physical CAN devices is feasible based on input data obtained from a simulation environment to test the CAN system of SDCs.
\tool provides the technological possibility of testing ECUs individually focusing on CAN messages as input and output.
Instead of manually testing CAN devices by defining specific CAN messages upfront, which is the standard industrial approach of our evaluation partner \aicas, using simulated signals sent over the CAN bus enables more realistic input data for the CAN devices since the CAN messages are based on simulated scenarios.
Furthermore, with the modular architecture of \tool, regression testing for SDCs can be enabled in co-simulation environments by implementing the given APIs.
This enables future research on SDC testing involving co-simulation environments.

The \tool component extends the \scissor tool by supporting data transmission over the CAN bus to test CAN devices.
We showed the usefulness of the tool in practice by integrating it into the DevOps pipeline of \aicas.
The tool enables regression testing for SDC CAN devices based on signals generated from simulators such as \beamng or \carla.
We believe that \tool enables future research on testing CAN devices and SDCs in general based on state-of-the-art simulation data.

\section*{Acknowledgements}
We thank the Horizon 2020 (EU Commission) support for the project COSMOS (DevOps for Complex Cyber-physical Systems), Project No. 957254-COSMOS) and the DFG project STUNT (DFG Grant Agreement n. FR 2955/4-1).

\balance
\bibliographystyle{IEEEtran}
\bibliography{main.bib}

\end{document}